\documentclass[conference]{IEEEtran}
\usepackage{cite, color}
\usepackage[normalem]{ulem}
\usepackage{longtable}
\usepackage[acronym,nonumberlist,nogroupskip,style=super]{glossaries}
\usepackage{array}
\usepackage{booktabs}
\usepackage{multirow}
\ifCLASSINFOpdf
   \usepackage [pdftex]{graphicx}
   \DeclareGraphicsExtensions{.pdf,.jpg,.png}
\else
   \usepackage [dvips]{graphicx}
\DeclareGraphicsExtensions{.eps}
\fi
\usepackage{amsmath}
\usepackage{amssymb}
\usepackage{enumitem}
\usepackage[cmintegrals]{newtxmath}
\ifCLASSOPTIONcompsoc
  \usepackage[caption=false,font=normalsize,labelfont=sf,textfont=sf]{subfig}
\else
  \usepackage[caption=false,font=footnotesize]{subfig}
\fi
\usepackage{url}
\newcommand{\rulesep}{\unskip\ \vrule\ }

\newacronym{e2e}{E2E}{End to End}
\newacronym{m2m}{M2M}{Machine-to-Machine}
\newacronym{iot}{IoT}{Internet of Things}
\newacronym{rpl}{RPL}{Routing Protocol for Low Power and Lossy Networks}
\newacronym{ids}{IDS}{Intrusion Detection System}
\newacronym{rfid}{RFID}{Radio Frequency Identification}
\newacronym{wsn}{WSN}{Wireless Sensor Network}
\newacronym{6lowpan}{6LoWPAN}{IPv6 over Low-powered Wireless Personal Area Network}
\newacronym{ietf}{IETF}{Internet Engineering Task Force}
\newacronym{ipv6}{IPv6}{Internet Protocol version 6}
\newacronym{tcp}{TCP}{Transmission Control Protocol}
\newacronym{of}{OF}{Objective Function}
\newacronym{dag}{DAG}{Directed Acyclic Graph}
\newacronym{dodag}{DODAG}{Destination Oriented Directed Acyclic Graph}
\newacronym{ch}{CH}{Cluster Head}
\newacronym{roll}{ROLL}{Routing Over Low-power and Lossy Networks working group}
\newacronym{of0}{OF0}{Objective Function Zero}
\newacronym{mrhof}{MRHOF}{Minimum Rank with Hysteresis Objective Function}
\newacronym{ext}{EXT}{Expected Transmission Count}
\newacronym{extof}{EXTOF}{Expected Transmission Count Objective Function}
\newacronym{lbof}{LBOF}{Load Balancing Objective Function}
\newacronym{cnc}{CNC}{Child Node Count}
\newacronym{taof}{TAOF}{Traffic Aware Objective Function}
\newacronym{ptr}{PTR}{Packet Transmission Rate}
\newacronym{dodagid}{DODAGID}{DODAG Identification}
\newacronym{dio}{DIO}{DODAG Information Object}
\newacronym{dis}{DIS}{DODAG Information Solicitation}
\newacronym{dao}{DAO}{Destination Advertisement Object}
\newacronym{daoack}{DAO-ACK}{DAO Acknowledgement}
\newacronym{P2P}{P2P}{Point to Point communication}
\newacronym{P2MP}{P2MP}{Point to Multi-Point communication}
\newacronym{MP2P}{MP2P}{Multi-Point to Point communication}
\newacronym{cc}{CC}{Consistency Check}
\newacronym{icmp}{ICMP}{Internet Control Messgae Protocol}
\newacronym{ccm}{CCM}{Counter wih CBC-MAC "Cipher Block Chaining - Message Authentication Code"}
\newacronym{aes}{AES}{Advanced Encryption Standard}
\newacronym{macl}{MAC}{Media Access Control}
\newacronym{macc}{MAC}{Message Authentication Code}
\newacronym{rsa}{RSA}{Rivest-Shamir-Adleman encryption}
\newacronym{sha}{SHA}{Secure Hash Algorithm}
\newacronym{os}{OS}{Operating System}
\newacronym{gps}{GPS}{Global Positioning System}
\newacronym{rtt}{RTT}{Round-Trip Time}
\newacronym{phy}{PHY}{Physical layer}
\newacronym{coap}{CoAP}{Constrained Application Protocol}
\newacronym{ipsec}{IPSec}{Internet Protocol Security}
\newacronym{dtls}{DTLS}{Datagram Transport Layer Security protocol}
\newacronym{rssi}{RSSI}{Received Signal Stregnth Indicator}
\newacronym{rss}{RSS}{Received Signal Stregnth}
\newacronym{dos}{DoS}{Denial of Service}
\newacronym{esp}{ESP}{Encapsulated Security Header}
\newacronym{tpm}{TPM}{Trusted Platform Module}
\newacronym{ernt}{ERNT}{Extended RPL Node Trustworthiness}
\newacronym{tof}{TOF}{Trust Objective Function}
\newacronym{srpl}{SRPL}{Secure RPL}
\newacronym{dht}{DHT}{Distributed Hash Table}
\newacronym{trail}{TRAIL}{Trust Anchor Interconnection Loop}
\newacronym{vera}{VeRA}{Version attack and Rank Authentication}
\newacronym{mop}{MOP}{Mode of Operation}
\newacronym{mop2}{MOP2}{mode of operation}
\newacronym{sprt}{SPRT}{Sequential Probability Ratio Test}
\newacronym{qos}{QoS}{Quality of Service}
\newacronym{ami}{AMI}{Advanced Metering Infrastructure}
\newacronym{cr}{CR}{Cognitive Radio}
\newacronym{loadng}{LOADng}{Low power and Lossy Networks On-demand Ad-hoc Distance-vector routing protocol - Next Generation}
\newacronym{loadng-ctp}{LOADng-CTP}{LOADng with Collection Tree Protocol}
\newacronym{corpl}{CORPL}{Cognitive and Opportunistic RPL}
\newacronym{aodv}{AODV}{Ad-hoc On-demand Distance Vector}
\newacronym{pdr}{PDR}{packet delivery rate}

\makeglossaries
\glsaddall

\begin{document}
%
\title{Secure Routing in IoT: Evaluation of RPL's Secure Mode under Attacks}

\author{\IEEEauthorblockN{Ahmed~Raoof}
\IEEEauthorblockA{Dep. of Systems and Computer Eng.\\
Carleton University, Canada\\
Email: ahmed.raoof@carleton.ca}
\and
\IEEEauthorblockN{Ashraf~Matrawy}
\IEEEauthorblockA{School of Information Technology\\
Carleton University, Canada\\
Email: ashraf.matrawy@carleton.ca}
\and
\IEEEauthorblockN{Chung-Horng~Lung}
\IEEEauthorblockA{Dep. of Systems and Computer Eng.\\
Carleton University, Canada\\
Email: chlung@sce.carleton.ca}
}

\maketitle

\begin{abstract}
As the Routing Protocol for Low Power and Lossy Networks (RPL) became the standard for routing in the Internet of Things (IoT) networks, many researchers had investigated the security aspects of this protocol. However, no work (to the best of our knowledge) has investigated the use of the security mechanisms included in the protocol's standard, due to the fact that there was no implementation for these features in any IoT operating system yet. A partial implementation of RPL's security mechanisms was presented recently for Contiki operating system (by Perazzo \textit{et al.}), which provided us with the opportunity to examine RPL's security mechanisms. In this paper, we investigate the effects and challenges of using RPL's security mechanisms under common routing attacks. First, a comparison of RPL's performance, with and without its security mechanisms, under three routing attacks (Blackhole, Selective-Forward, and Neighbor attacks) is conducted using several metrics (e.g., average data packet delivery rate, average data packet delay, average power consumption... etc.) Based on the observations from this comparison, we come up with few suggestions that could reduce the effects of such attacks, without having added security mechanisms for RPL.
\end{abstract}

\IEEEpeerreviewmaketitle

\section{Introduction}
Routing is one of the most researched fields in the world of \gls{iot}, due to the constraint nature of these devices. Introduced by \gls{ietf}, the \gls{rpl} \cite{RFC6550} had become the standard for routing in many \gls{iot} networks as it was designed to efficiently use the constraint resources of \gls{iot} devices while providing effective routing service. Routing security was an integral part of \gls{rpl}'s design with several, but optional, security mechanisms available\cite{RFC6550}.

Since it became a standard in 2012, \gls{rpl} gained a lot of research interest, with many of the literature focusing on the security aspects of routing using the protocol, such as: types of routing attacks, new mitigation methods and \glspl{ids}, and security-minded \glspl{of}\cite{TAOF,Djedjig2015,Djedjig2017, Karkazis2014,Airehrour2019}. Interestingly, there was no research discussing the effects of using \gls{rpl}'s security mechanisms, specifically under routing attacks. This is most probably due to the lack of implementation of \gls{rpl}'s security mechanisms in any of the available \gls{iot} \gls{os}, such as Contiki \gls{os}\cite{ContikiOSRef} and TinyOS \cite{TinyOS}.

However, recently Perazzo \textit{et al.} in \cite{Perazzo2017} provided a partial implementation of \gls{rpl}'s security mechanisms for Contiki OS, which added the Preinstalled secure mode and the optional replay protection mechanism. This implementation provided us with the basis upon which the work in this paper is built on. In this paper, we have experimentally investigated \gls{rpl}'s performance under three common routing attacks using several metrics to analyze and compare the performance between having RPL's security mechanisms enabled or disabled.

Our contributions can be summarized in the following points: \textbf{(1)} We provided a performance comparison for RPL between the unsecure mode and the Preinstalled secure mode; the latter case is examined with and without the optional replay protection. We discovered that running RPL in the Preinstalled secure mode (without replay protection) does not use more resources than the unsecure mode, even under attack. \textbf{(2)} We verified that the Preinstalled secure mode is able to stop external adversaries from joining the IoT network for the investigated attacks. Further, We showed that the optional replay protection also provides an excellent mitigation against the Neighbor attack; however, it needs further optimization to reduce its effect on energy consumption. and \textbf{(3)} We observed and analyzed the effect of the investigated attacks on the routing topology and proposed a few simple techniques that could help reduce the effects of the investigated attacks, without using external security measures such as IDSs or added security mechanisms.


The rest of this paper goes as follows: Section \ref{rltdwrk} looks into the related works. In section \ref{background} an overview of \gls{rpl} and its security mechanisms is presented. Section \ref{RPLEval} discusses our evaluation methodology, setup, assumptions, adversary model, and attack scenarios. Results from the evaluation are shown in section \ref{resultsanalysis}. Section \ref{discussionSec} discuses our observations from the results and few suggestions we are proposing to be used when designing \gls{rpl}-based \gls{iot} networks.

\section{Related Works}\label{rltdwrk}
In this section, a highlight on some of the influencing literature that discussed \gls{rpl}'s performance under common routing attacks is presented. As we mentioned earlier, none of them had investigated \gls{rpl}'s security mechanisms.

%
Le \textit{et al.} in \cite{Le2013a} evaluated \gls{rpl}'s performance under four \gls{rpl}-based attacks: the Decreased Rank attack, Local Repair attack, Neighbor attack, and DODAG\footnote{DODAG = Destination-Oriented Directed Acyclic Graph} Information Solicitation (DIS) attack. Their work showed that the Decreased Rank attack and the Local Repair attack affects the \gls{pdr} the most, while \acrshort{dis} attack introduced the most \gls{e2e} latency. Neighbor attack showed the least impact on the network. Compared to our work, the authors only tackled with the unsecured mode of \gls{rpl}, and they didn't investigate the effect of their attacks on power consumptions.

Kumar \textit{et al.} in \cite{Kumar2016} investigated the effects of the Blackhole attack, on \gls{rpl}-based network through simulations. As expected, the attack was successful in reducing the \gls{pdr} and increased both the \gls{e2e} latency and control messages overhead. However, the authors did not evaluate the power consumption and neglected the existence of \gls{rpl}'s security mechanisms.

Perazzo \textit{et al.} in \cite{Perazzo2017} provided the first, standard-compliant as per their claim, partial implementation of \gls{rpl} security mechanisms. One secure mode, the Preinstalled secure mode, and the optional replay protection, named the \gls{cc} mechanism, were introduced to ContikiRPL (Contiki OS version of \gls{rpl}). The authors provided an evaluation for their implementation, and compared \gls{rpl}'s performance between using and not using the Preinstalled secure mode. However, It is worth noting that the authors did not evaluate their implementation against actual attacks.

Our previous work in \cite{Raoof2019} presented the first glimpse of the effect that \gls{rpl}'s security mechanisms could have on \gls{rpl}-based \gls{iot} networks when there is an actual attack. \gls{rpl}'s performance (with and without the preinstalled secure mode) was investigated under three attacks: the Blackhole, Selective-Forward, and Neighbor attacks using simulations. The preliminary results showed that \gls{rpl}'s secure modes could mitigate the external adversaries of the investigated attacks, but not the internal adversaries. However, as it is an ongoing work, we were not able to provide deeper analysis on the results, nor to inspect the optional replay protection mechanism.

\section{Background Review}\label{background}

\subsection{RPL Overview} \label{RPLOverview}
\gls{rpl} was developed as a distance-vector routing protocol\cite{RFC6550}. It arranges the network devices into a \glspl{dodag}\cite{Janicijevic2011}: a network of nodes connected without loops and where the traffic is directed toward one \textit{root} or \textit{sink} node\cite{RFC6550,Granjal}. The creation of the \gls{dodag} depends on the used \textit{\gls{of}}, which defines essential configurations such as the used routing metrics, how to calculate the \textit{rank} (the rank of a node represents its distance to the root node based on the routing metrics defined by the \gls{of}), and how to select parents in the \gls{dodag}. To accommodate the different applications and environments where \gls{rpl} can be deployed, \gls{rpl} has several \glspl{of}\cite{RFC6552, RFC6719, TAOF} available for use\cite{Raoof2018}. Also, deployments of \gls{rpl} can have their own \glspl{of}.

Three types of traffic are supported by \gls{rpl}: \gls{MP2P} traffic (nodes to sink) through normal \gls{dodag}, \gls{P2MP} traffic (sink to nodes) through source routing, and \gls{P2P} traffic (non-root node to non-root node) through \gls{rpl}'s \textit{Modes of Operation (MOP)}\cite{RFC6550}, which dictate how the downward routes are created. 

\gls{rpl} has five types of control messages; four of them have two versions (base and secure versions), and the last one has only a secure version. The secure version of \gls{rpl}'s control messages adds new unencrypted header fields and either a \gls{macc} or a digital signature field to the end of the base version, then encrypts the base part and the \gls{macc}\cite{RFC6550}.

\gls{dio} and \acrfull{dis} messages are used for the creation and maintenance of the \gls{dodag}\cite{RFC6550}. The root node starts the \gls{dodag} creation by multicasting a \gls{dio} message that contains the essential \gls{dodag} configurations and the root node's rank (the root node has the lowest rank in the \gls{dodag}). Upon receiving a \gls{dio} message, each node will select its \textit{preferred parent}, calculate its own rank, and multicast a new \gls{dio} with its calculated rank\cite{RFC6550, Raoof2018}. \acrshort{dis} messages are used to solicit \gls{dio} messages from node's neighbors when it is needed, e.g., a new node wants to join the networks or no \gls{dio} messages had arrived for a long time\cite{RFC6550}.

\gls{dao} and \glspl{daoack} messages are the backbones of the downward routes creation\cite{RFC6550}. The \gls{dao} contains path information about reachable nodes by its sender, and depending on \gls{rpl}'s \acrlong{mop2} it will be used to create the downward routing table. Based to the \gls{dodag}'s configurations, a flag in \gls{dao} message will mandate an acknowledgment (\gls{daoack} message) from the receiver.

\subsection{RPL's Security Mechanisms}
To secure the routing service, RPL either relies on the security measures at Link layer (i.e. IEEE 802.15.4\cite{STND802154}), or uses its own security mechanisms, resembled in three modes of security and an optional replay protection mechanism\cite{RFC6550, Perazzo2017}: 
The default mode for RPL is the \textit{\textbf{Unsecured}} mode (UM), where only link-layer security is applied, if available. The second mode, the \textit{\textbf{Preinstalled}} secure mode (PSM), which uses the preinstalled symmetrical encryption keys to secure RPL control messages. Finally, the \textit{\textbf{Authenticated}} security mode (ASM) uses the preinstalled keys for the nodes to join the network; after which all routing-capable nodes have to acquire new keys from an authentication authority. To protect the routing service from replay attacks, RPL uses Consistency Checks as an optional mechanism that can be used with either the preinstalled (PSMrp) or authenticated mode (ASMrp). In these checks, a special secure control message (\gls{cc} message) with non-repetitive nonce value are exchanged and used to assure no replay had occurred\cite{RFC6550}.

\section{Evaluation of RPL's Security Mechanisms under Attacks}\label{RPLEval}
In this paper, \gls{rpl} performance is evaluated against three attacks\cite{Raoof2018, Wallgren2013}: the Blackhole, the Selective-Forward, and the Neighbor attacks. Experiments were conducted with \gls{rpl} in the unsecure mode (vanilla ContikiRPL) and the Preinstalled secure mode (as in Perazzo \textit{et al.}\cite{Perazzo2017} implementation). For the latter, we evaluated \gls{rpl} with and without the optional replay protection mechanism.

\subsection{Evaluation Setup}\label{EvalSetup}
\begin{figure}[!t]
\centering
\includegraphics[scale=0.41]{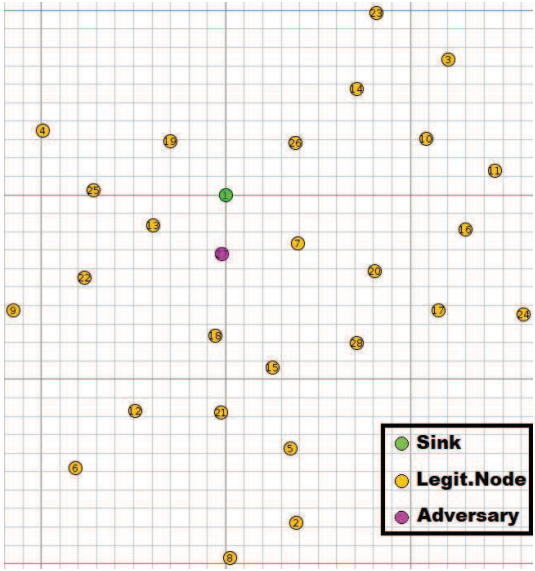}
\caption{Network topology (better viewed in colors.)}
\label{fig_1}
\end{figure}

Cooja, the simulator for Contiki OS\cite{ContikiOSRef}, was used for all the simulations (with simulated motes). Fig.\ref{fig_1} shows the topology used in our evaluation. A list of simulation parameters is provided in Table \ref{table_1}.

\begin{table}[!t]
	\caption{List of Simulation Parameters}
	\label{table_1}
	\begin{tabular}{ll}
		\toprule
		\textbf{Description}&\textbf{Value}\\
		\midrule
		No. of experiments&
		\begin{minipage}{0.38\columnwidth}
			Four (See \S\ref{AttModANDSen})
		\end{minipage} \\
		\midrule
		No. of scenarios per experiment&4 scenarios\\
		\midrule
		No. of sim. rounds per scenario / time&10 rounds / 20 min. per round\\
		\midrule
		Node Positioning&Random\\
		\midrule
		Deployment area&290m W x 310m L\\
		\midrule
		Number of nodes&28 (adversary included)\\
		\midrule
		Sensor nodes type&
		\begin{minipage}{0.38\columnwidth}
			Arago Sys. Wismote mote.
		\end{minipage}\\
		\midrule
		Propagation model&
		\begin{minipage}{0.38\columnwidth}
			Unit Disk Graph Model
		\end{minipage}\\
		\midrule
		DATA transmission rate& $\simeq$ 1 packet per minute\\
		\bottomrule
	\end{tabular}
\end{table}

Our topology represents a single \gls{dodag} network that has one root or sink node (green node). To reduce the complexity of the observed metrics,  only one adversary (purple node) was used for the attacks. This adversary was positioned near the sink node, as that would introduce the biggest effect of the investigated attacks\cite{Pu2018a,Wallgren2013,Mayzaud2016}. The targeted nodes for the attacks are (2, 5, 6, 8, 12, 15, 18, 21, and 28), with node (28) providing an alternative path for the targeted nodes to send their packets toward the sink. Having an alternative path is crucial to our experiments to examine how will the self-healing mechanisms of RPL respond to the attacks.

Note that we tried to implement the simulations using Zolertia Z1 motes\cite{Z1} (has 8KB RAM and 92KB Flash memory) to compare our results to \cite{Raoof2019}. However, enabling the replay protection mechanism of \gls{rpl} in our simulation caused the mote to always run out of RAM, rendering the simulation impractical. Hence, we moved to the more powerful Wismote mote (has 16KB RAM and a 256KB Flash memory\cite{Wismote}).

\subsection{Assumptions}\label{EvalAssump}
The following assumptions were used in our evaluation: all legitimate nodes are sending one data packet per minute toward the sink, while the adversary only participates in the \gls{dodag} formation without sending any data packets. \gls{rpl} is using the default \gls{of}, namely the \gls{of0}\cite{RFC6552}. To keep the focus on \gls{rpl} at the Network layer, we assumed no security measure was enabled at the Link layer. In other words,  no encryption was used at the Link layer. All the attacks were also implemented at the Network layer.

The results obtained from the simulations were averaged over ten rounds for each scenario with a 95\% confidence level.

\subsection{Adversary Model and Attack Scenarios}\label{AttModANDSen}
We conducted four experiments: the first three experiments (RPL in UM, RPL in PSM, and RPL in PSMrp) have an \textit{internal} adversary; who participates in the creation of the topology from the beginning (and has the preinstalled encryption keys in the 2$^{nd}$ and 3$^{rd}$ experiments), and the fourth experiment (RPL in PSM) uses an \textit{external} adversary who runs RPL in the UM and does not have knowledge of the secure versions of \gls{rpl}'s control messages, while the legitimate nodes run RPL in the PSM. Table \ref{Exp_Sum} lists the settings for these experiments.

\begin{table}[!t]
	\caption{Experiments summary}
	\label{Exp_Sum}
	\centering
	\begin{tabular}{c|c|c|c}
		\toprule
		\textbf{Experiment} & \textbf{Secure Mode} & \textbf{Replay Protection} & \textbf{Adversary Type}\\
		\toprule
		UM-I & $\times$ & $\times$ & Internal (I)\\
		\midrule
		PSM-I & \checkmark & $\times$ & Internal (I)\\
		\midrule
		PSMrp-I & \checkmark & \checkmark & Internal (I)\\
		\midrule
		PSM-E & \checkmark & $\times$ & External (E)\\
		\bottomrule
	\end{tabular}
\end{table}

The adversary will always start as a legitimate node, try to join the network and actively participate in the creation and maintenance of the \gls{dodag}, work as a legitimate node for two minutes (to assure full integration with the network), then it will launch the attack afterward.

For the attacks themselves, we have four scenarios: (i) \textit{No Attack}: the adversary works as a fully legitimate node, (ii) the \textit{Blackhole Attack}: the adversary drops all the traffic coming through (\gls{rpl} control messages and Data Packets) \cite{Raoof2018}, (iii) the \textit{Selective-Forward Attack (SF)}: the adversary drops data packets only (\gls{rpl} control messages will pass normally) \cite{Pu2018a}, and (iv) the \textit{Neighbor Attack}: the adversary would pass any \gls{dio} message it receives from its neighbors without any processing or modification \cite{Le2013a}.

The choice of these three attacks was based on the fact they have the minimum cost for the adversary to launch them, as they require very little or no processing of RPL’s messages. At the same time the effect of these attacks can be significant on the network.

\section{Results and Analysis}\label{resultsanalysis}
\begin{figure*}[!t]
\centering
\subfloat[Average packet delivery rate (\gls{pdr}).]{\includegraphics[scale=.32]{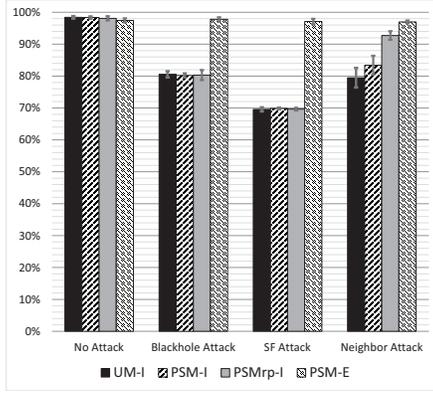}%
\label{fig_2a}}
\hfil
\subfloat[Average network \gls{e2e} latency.]{\includegraphics[scale=.32]{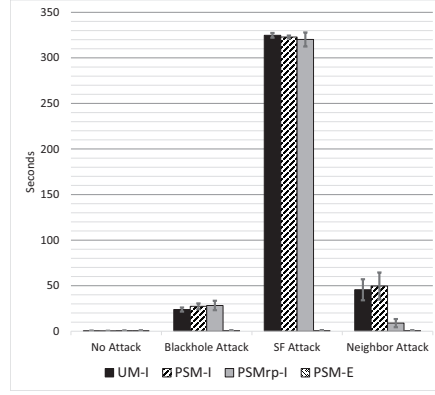}%
\label{fig_2b}}
\hfil
\subfloat[Exchanged \gls{rpl} control messages.]{\includegraphics[scale=.32]{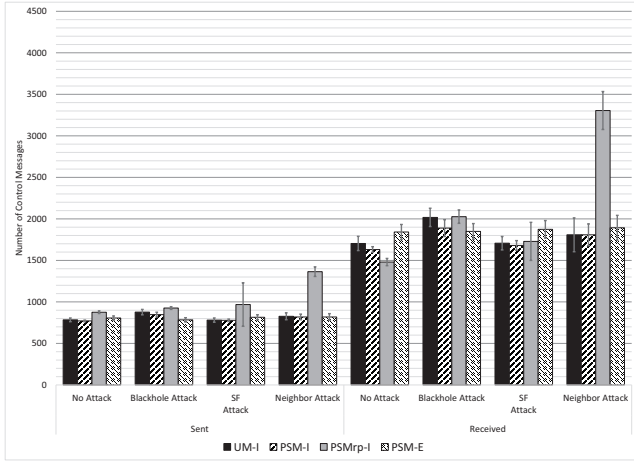}%
\label{fig_2c}}
\hfil
\subfloat[Average network power consumption, per received packet.]{\includegraphics[scale=.32]{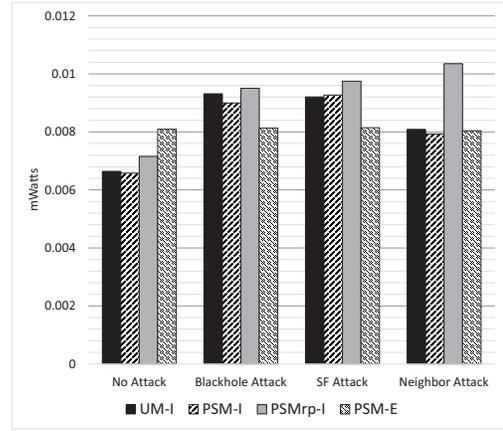}%
\label{fig_2d}}
\hfil
\caption{Simulation results for the four experiments. (UM-I: unsecure mode-internal	adversary, PSM-I: preinstalled secure mode-internal adversary, PSMrp-I: preinstalled secure mode with replay protection-internal adversary, PSM-E: preinstalled secure mode-external adversary.)}
\label{fig_2}
\end{figure*}

\begin{figure*}[ht]
	\centering
	\subfloat[No Attack scenario and SF Attack scenario (all experiments except PSM-E)]{\includegraphics[scale=.256]{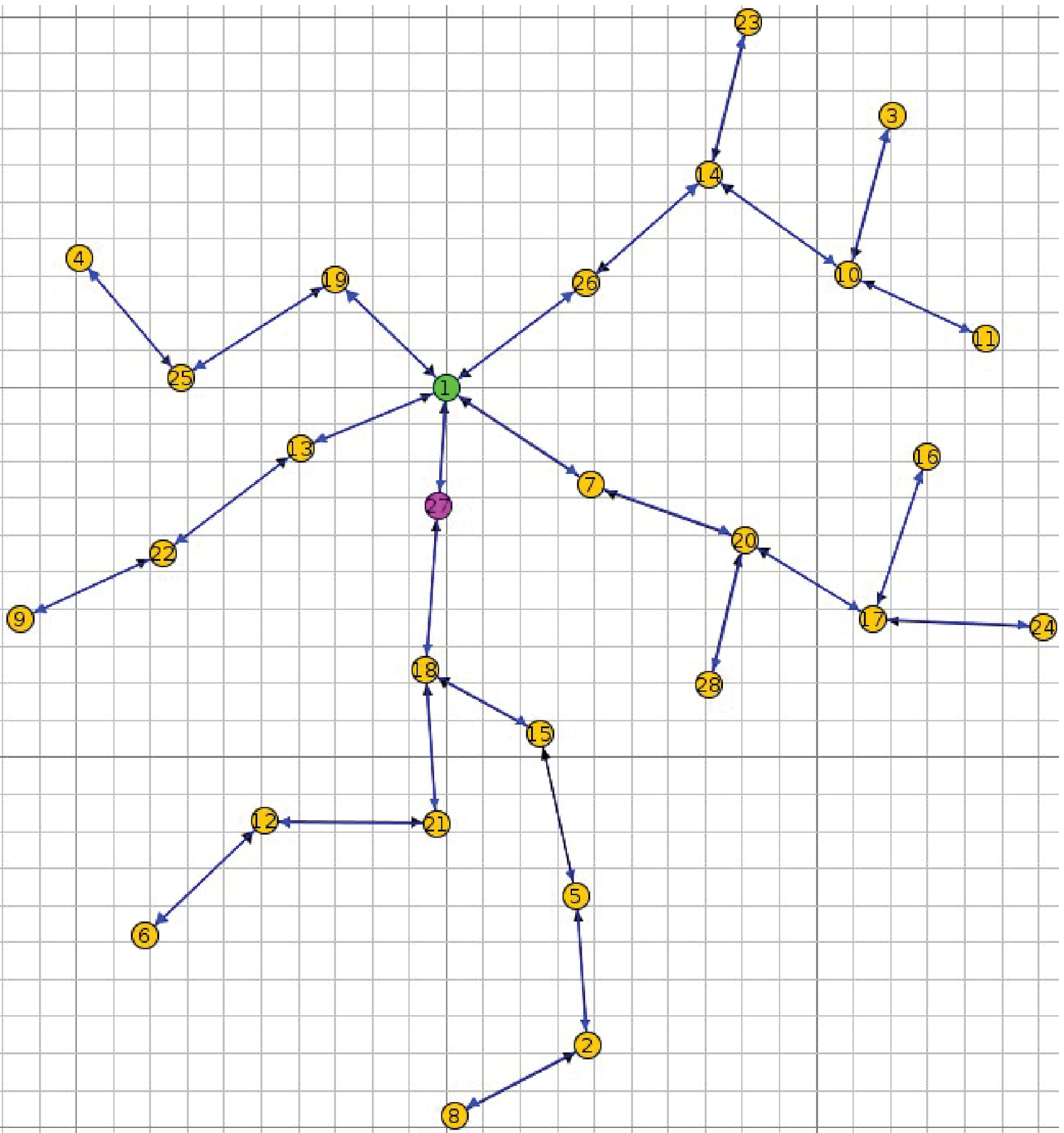}%
		\label{fig_3a}}
	\rulesep
	\subfloat[Blackhole Attack scenario (all experiments), and all scenarios for PSM-E.]{\includegraphics[scale=.25]{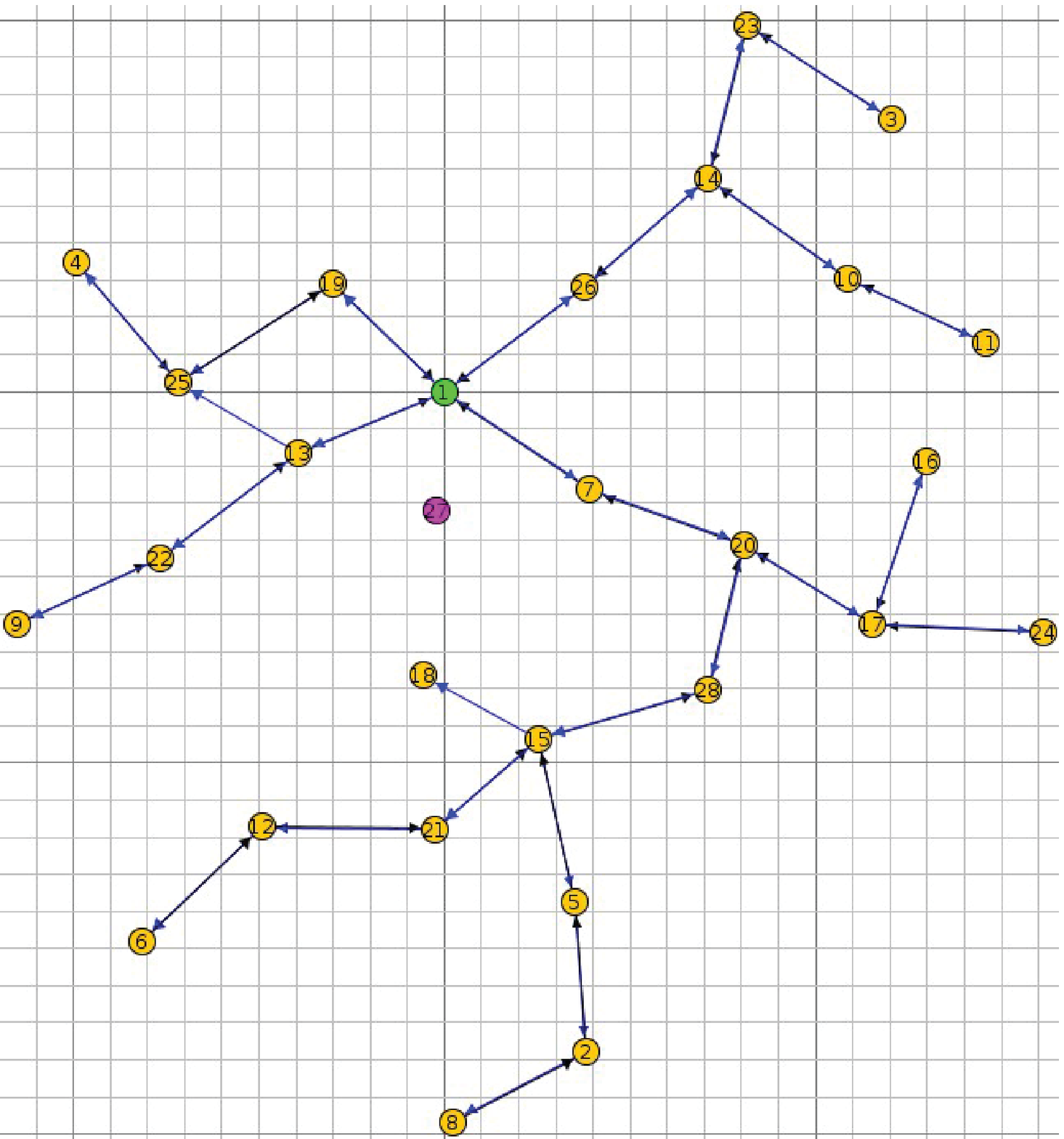}%
		\label{fig_3b}}
	\rulesep
	\subfloat[Neighbor Attack scenario (UM-I and PSM-I).]{\includegraphics[scale=.25]{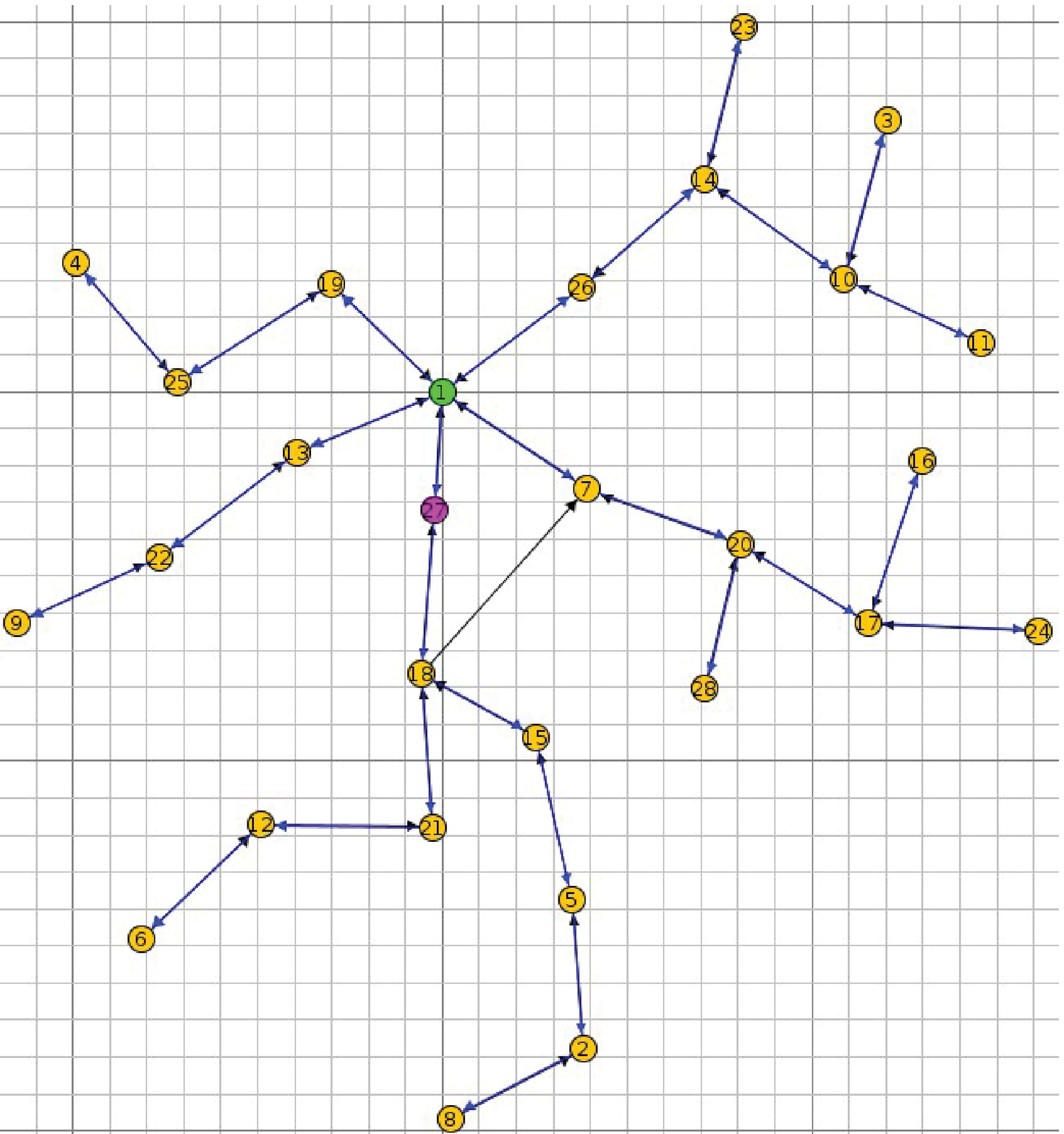}%
		\label{fig_3c}}
	\rulesep
	\subfloat[Neighbor Attack scenario (PSMrp-I).]{\includegraphics[scale=.25]{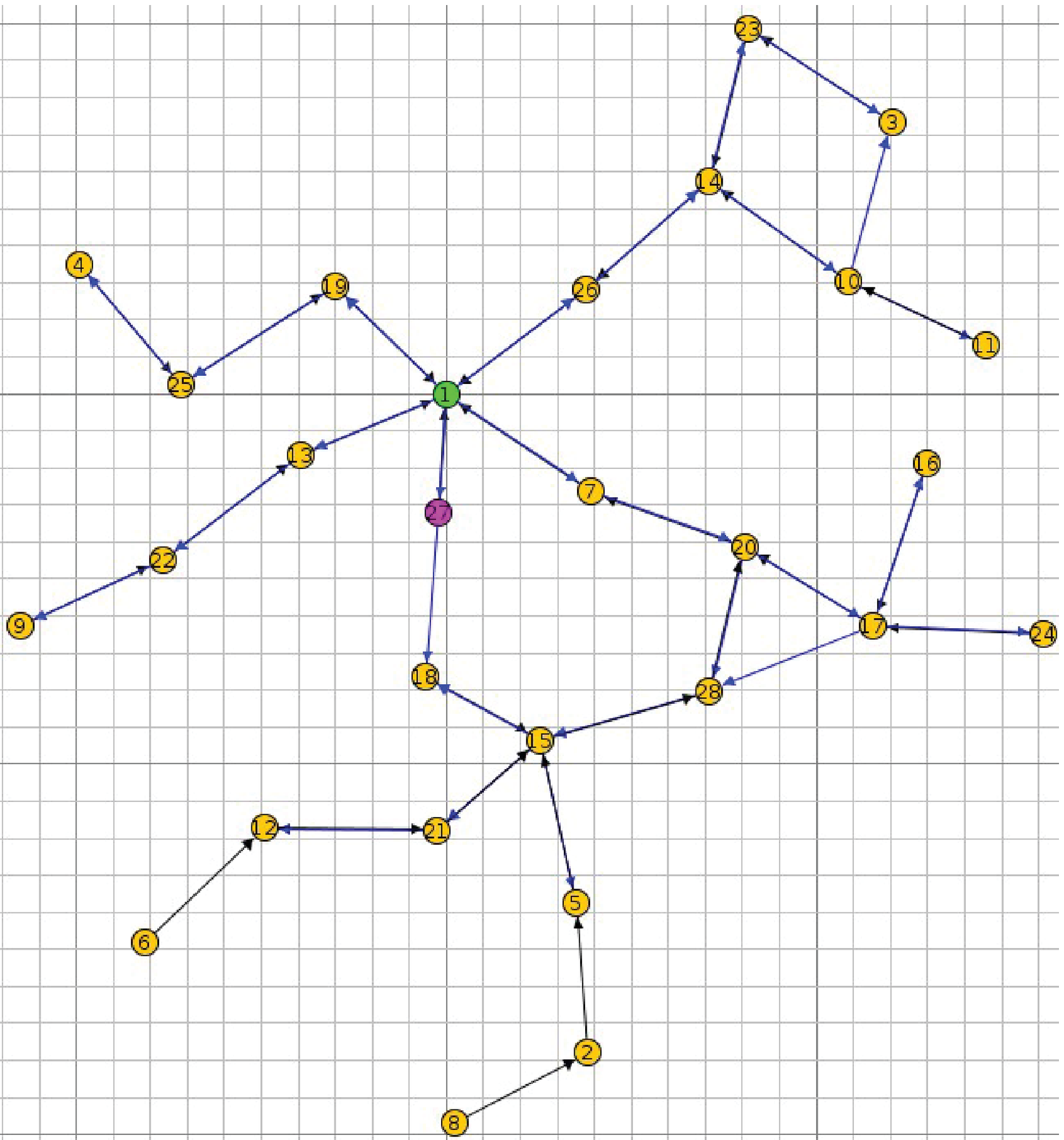}%
		\label{fig_3d}}
	\caption{Routing \glspl{dodag} during the investigated scenarios.}
	\label{fig_3}
\end{figure*}

Fig.\ref{fig_2} shows the results from all the experiments expressed as the average \gls{pdr}, average \gls{e2e} latency, the number of exchanged \gls{rpl} control messages, and average network power consumption (per received packet). In addition, Fig.\ref{fig_3} presents the routing \gls{dodag} for each scenario that was formed in 90\% of the time in all four experiments.

\textbf{Effects on \acrfull{pdr}:} Looking at Fig.\ref{fig_2a}, It is clear that the Preinstalled secure mode of RPL was able to successfully mitigate the three attacks when the adversary is external with the \gls{pdr} hovering around 98\%.

On the other hand, when the adversary is internal, the SF attack has the most effect (in all three experiments) on the \gls{pdr}, decreasing it to a low of 70\%. The main reason behind the success is that the adversary, due to being an active participant in the DODAG maintenance, is always chosen as the preferred parent for its sub-\gls{dodag} but none of their data packets are passed to the sink node. Fig.\ref{fig_3a} shows the network \gls{dodag} during the SF attack.

For the Blackhole attack, the self-healing mechanisms of RPL were always able to detect the unresponsive adversary after some time (in our experiments, self-healing kicked in after about ten minutes from the attack launch time) and initiated a local repair for the affected sub-\gls{dodag} to switch to an alternative path. Hence, not all data packets got dropped, and that explains why the \gls{pdr} was in the range of 80\%. Fig.\ref{fig_3b} shows how the \gls{dodag} isolated the adversary and selected the alternative path after ten minutes from the attack launch time.

Finally, for the Neighbor attack, the adversary was able to reduce the \gls{pdr} for the UM-I and PSM-I experiments, as node 18 always chooses either node 7 or 13 as its preferred parent (Fig.\ref{fig_3c} shows the node 18 selecting node 7 as its preferred parent), due to receiving their \gls{dio} messages through the adversary. Since nodes 7 and 13 are actually out of node 18's range, all packets sent toward them from node 18 and its sub-\gls{dodag} are lost. Hence, the \gls{pdr} is in the same range as in the Blackhole attack scenario. However, activating the replay protection mechanism results in much better \gls{pdr} as the mechanism verifies the original sender of each \gls{dio} message before processing its contents. Fig.\ref{fig_3d} demonstrate how the network (in PSMrp-I experiment) opted for the alternative path after few minutes from launching the Neighbor attack.

\textbf{Effects on the \gls{e2e} latency:} Confirming our findings mentioned above, Fig.\ref{fig_2b} shows that the RPL's Preinstalled secure mode mitigated the three attacks when they are launched by an external adversary, keeping the \gls{e2e} latency at minimum.

Due to the large number of undelivered data packets for the affected nodes, the SF had the largest \gls{e2e} latency among all the internal attacks. This effect is, again, because of the active participation of the adversary in the \gls{dodag} maintenance.

For the same reason, the Blackhole attack introduced some latency to the network. However, since the affected nodes were able to find an alternative path and were successful in delivering the rest of their data packets, the latency was much less than in the case of the SF attack scenario.

The situation is more complicated for the Neighbor attack scenario, as self-healing mechanisms were used several times to recover the affected nodes from the attack, which led to even higher \gls{e2e} latency than the Blackhole attack scenario. In general, whenever the node 18 switches its preferred parent to node 7 or 13, the sub-\gls{dodag} suffers from Blackhole-like conditions resulting in losing several data packets. In addition, node 18 will either switch its preferred parent back to the adversary when it does not receive \gls{dio} messages from the "ghost parent" (node 7 or node 13), or initiate a local repair procedure (if \gls{dodag} inconsistencies were detected) that results in the whole sub-\gls{dodag} choosing the alternative path to deliver their packets. Either way it will add more latency to the network.
Using the replay protection will significantly reduce the latency from the Neighbor attack, as the node 18 will not switch its preferred parent as long as it does not receive the correct \gls{cc} response from nodes 7 and 13.

\textbf{Effects on the exchanged number of \gls{rpl}'s control messages:} As seen in Fig.\ref{fig_2c}, the number of control messages exchanged in the network is almost the same for all experiments and all the scenarios, with the replay protection mechanisms adding a bit more control messages. The exception of this conclusion is the Neighbor attack scenario with \gls{rpl} in the Preinstalled secure mode and the replay protection mechanism is active. In this special case, the replay protection mechanism introduced a much higher number of control messages, due to the exchange of the \gls{cc} messages whenever a "ghost" \gls{dio} message is received by nodes 7, 13, or 18.

It is worth noting that the number of received control messages is always higher than the sent one, because many of the sent control messages are multicast messages which will be received by all neighboring nodes of the sender.

\textbf{Effects on power consumption:} Fig.\ref{fig_2d} shows the average network power consumption per received packet, as it gives a more accurate look into the effect of the attacks on the power consumption than just using the regular average power consumption readings.

Looking at the results of the external adversary experiment in the No Attack scenario, we can see that power consumption is a bit higher compared to the same scenario in the other experiments. The reason is that the data packets from the affected nodes are taking the alternative and longer path, i.e., more power is used by the nodes on that path. However, the power consumption pattern is identical in all the scenarios of the external adversary experiment, which indicates it is not affected by the attacks; hence, a successful mitigation of the attacks.

For all the other experiments (with an internal adversary), the power consumption patterns (per each scenario) are very similar between the unsecure mode and the Preinstalled secure mode in the No Attack, Blackhole, and Selective-Forward attacks scenarios, with the replay protection mechanism having a bit more power consumption than the rest. This is due to the fact that many data packets were not delivered and the power consumed for their unsuccessful deliveries is fully wasted.

Now, it is clear from Fig.\ref{fig_2d} that using the replay protection significantly increases the average power consumption when the Neighbor attack is launched, even if almost all of the sent data packets were delivered successfully. This time the reason behind this behavior is the increased number of control messages exchanged to mitigate the attack, as seen in Fig.\ref{fig_2c}.

\section{Discussions}\label{discussionSec}
We hope the work and results in this paper will encourage more work evaluating RPL security evaluation and encourage research that confirms, updates, or extends our results.

\subsection{Observations}\label{observ}
Our observations from the results mentioned above and their analysis can be summarized in the following points:-
\begin{itemize}
	\item Using RPL's Preinstalled secure mode (and by extension, the Authenticated secure mode) can mitigate the external adversaries of the investigated attacks. However, a further investigation should be conducted using an external adversary who can operate \gls{rpl} in PSM but does not have the encryption keys.
	\item RPL's performance using RPL's Preinstalled secure mode (without the replay protection mechanism) is similar to that in the unsecure mode, but with the added benefit of mitigating the external attacks in the scenarios investigated in this paper.
	\item Enabling RPL's replay protection mechanism will significantly reduce the effect of Neighbor attacks on \gls{pdr} and \gls{e2e} latency. However, in its current implementation, it will increase the power consumption as well, which can lead to energy depletion of the devices. In theory, an adversary can replay \gls{dio} messages regularly to keep the affected nodes always busy with the consistency checks, leading to depletion of their energy and to shutdown.
	\item RPL's secure modes require more memory and storage spaces than the unsecure mode, which means not all \gls{iot} devices can use them -- see \S\ref{EvalSetup}.
\end{itemize}

\subsection{Suggestions to Reduce the Effects of Routing Attacks on RPL's Performance}\label{Recomm}
Based on the observations mentioned above, we propose the following suggestions to help reduce the effects of routing attacks on RPL's performance. These suggestions do not require extra security mechanisms or systems. But, their effectiveness needs to be verified through more experiments:-
\begin{enumerate}
	\item Designing the network topology in a way where there are more alternative paths toward the root node and more neighbors per node. This would decrease the recovery time required for nodes to overcome a Blackhole attack and reduce the effects from the other investigated attacks on \gls{pdr} and \gls{e2e} latency.
	\item Optimizing the "dead parent" timeout of RPL to go with the network's changing conditions could decrease the \gls{e2e} latency and increase the \gls{pdr}. However, that may increase the power consumption when there is no attacks. We would recommend using a dynamic approach where the "Dead parent" timeout is randomized, or to use the \acrlong{6lowpan}-Neighbor Discovery (\acrshort{6lowpan}-ND) protocol\cite{RFC6775,RFC8505,ietf-6lo-ap-nd-12}, which works alongside RPL to detect node's neighbors and check their status in a resource-friendly way.
\end{enumerate}


\section*{Acknowledgment}
The second and the third authors acknowledge support from the Natural Sciences and Engineering Research Council of Canada (NSERC) through the Discovery Grant program.

\bibliographystyle{IEEEtran}
\bibliography{ARaoof-GlobeCom2019-Submission}

\end{document}